\documentclass{INTERSPEECH2023}
\usepackage{indentfirst}
\usepackage{makecell}
\usepackage{CJKutf8}
\usepackage{color}
\newcommand{\setParDis}{\setlength {\parskip} {0.4cm} }
\newcommand{\setParDef}{\setlength {\parskip} {0pt} }
\interspeechcameraready

\title{ComedicSpeech: Text To Speech For Stand-up Comedies \\ in Low-Resource Scenarios}

\name{Yuyue Wang$^{1,*}$, Huan Xiao$^{2,*}$~\thanks{$^{*}$  The first two authors contribute equally to this work.}, Yihan Wu$^{1}$, Ruihua Song$^{1,\#}$ \thanks{$^{\#}$ Corresponding author: songruihua\_bloon@outlook.com.}}

\address{
  $^{1}$ Gaoling School of Artificial Intelligence, Renmin University of China, China \\
  $^{2}$ School of Statistics, Renmin University of China, China 
  }

\email{\{wangyuyue123, xiaohuan001, yihanwu\}@ruc.edu.cn,
songruihua\_bloon@outlook.com}

\begin{document}

\maketitle

\begin{abstract}
Text to Speech (TTS) models can generate natural and high-quality speech, but it is not expressive enough when synthesizing speech with dramatic expressiveness, such as stand-up comedies. Considering comedians have diverse personal speech styles, including personal prosody, rhythm, and fillers, it requires real-world datasets and strong speech style modeling capabilities, which brings challenges. In this paper, we construct a new dataset and develop ComedicSpeech, a TTS system tailored for the stand-up comedy synthesis in low-resource scenarios. First, we extract prosody representation by the prosody encoder and condition it to the TTS model in a flexible way. Second, we enhance the personal rhythm modeling by a conditional duration predictor. Third, we model the personal fillers by introducing comedian-related special tokens. Experiments show that ComedicSpeech achieves better expressiveness than baselines with only ten-minute training data for each comedian. The audio samples are available at \url{https://xh621.github.io/stand-up-comedy-demo/}

\end{abstract}

\section{Introduction}
\indent

Many TTS models \cite{Tacotron, Tacotron2, fastspeech, fastspeech2} can synthesize natural and high-quality speech. However, they cannot perform well when synthesizing speech which requires dramatic expressive force, such as stand-up comedies. 
The stand-up comedy performance has a strong personal speech style, and its three characteristics bring great challenges to this task:
\begin{itemize}
    \item \textbf{Personal Prosody.} Stand-up comedians (also called ``comedians`` for short in our paper) always use various emotions, high or low intonations, and stress to enhance the expressive and humorous in their own styles, which demands stronger prosody modeling capability of the TTS system.
    \item \textbf{Personal Rhythm.} A personal rhythm related to content including variable pauses and speaking rates is always used by comedians to create humorous effects. These rhythmic variations show the comedian's personal performance style, requiring accurate phoneme duration predicting ability of the TTS system.
    \item \textbf{Personal Fillers~\footnote{In this paper, personal filler refers to a mantra, a habitual suffix, or a pause that occurs as personal language habits.}} Comedians often use personal fillers (abbreviated as PF) that do not affect the meaning of a sentence, e.g. ``well'', ``you know", ``I mean", but enhance the expressiveness of performance. In stand-up comedy scenario, PFs are highly speaker-dependent. Some of them are expressed as vague pronunciations with much faster speed rather than word by word, which makes it difficult to synthesize for TTS.    
\end{itemize}

\begin{table}[t!]
    \setlength{\abovecaptionskip}{0.05cm}
  \caption{An example of special token ($\left \langle spc 1 \right \rangle$) replacement. The Personal Filler (PF) which means ``you know'' is marked in red.}
  \renewcommand\arraystretch{0.8}
  \setlength{\tabcolsep}{0.8mm}{\label{tab: an example of replacement}
  \centering
  \begin{tabular}{ll}
    \toprule
    \makecell{raw text}                   & \begin{CJK}{UTF8}{gbsn} \makecell[ll]{我真正开始有一点自信以后，是我开始说脱\\口秀以后，我觉得我长得挺好的，\color{red}{你知}\color{red}{道吧}}                             \end{CJK} \\
    \midrule[0.5pt]
    \makecell{raw \\phoneme}                &   \makecell[ll]{uo3 zh e1 n zh e4 ng k ai1 sh ii3  iou3 i4 d ia3\\ n z ii4 x i4 n i3 h ou4 sh ii4 uo3 k ai1 sh ii3 sh\\ uo1 t uo1 k ou3 x iou4 i3 h ou4 uo3 j ve2 d e5\\ uo3 zh a3 ng d e2 t i3 ng h ao3 d e5  \color{red}{n i3 zh ii1}\\\color{red}{d ao4 b a5}}    \\
    \midrule
    \makecell{phoneme\\after \\ replacement}  &  \makecell[ll]{uo3 zh e1 n zh e4 ng k ai1 sh ii3 iou3 i4 d ia3\\ n z ii4 x i4 n i3 h ou4  sh ii4 uo3 k ai1 sh ii3 sh\\ uo1 t uo1 k ou3 x iou4 i3 h ou4 uo3 j ve2 d e5\\ uo3 zh a3 ng d e2 t i3 ng h ao3 d e5  \color{red}{$\left \langle spc 1 \right \rangle$}}                    \\
    \bottomrule
  \end{tabular}}
  
  \vspace{-2.0em}
\end{table}

\indent
Most previous works focus on improving the personal speech style by conditioning the source TTS model with the extracted personal representation~\cite{implicit1-gst, implicit2-agent, implicit3-content, implicit4-meta,expressive-context,adaspeech4}.
They introduce additional encoders, such as reference speech encoder, auxiliary text encoder, etc., to extract style from speech or predict style according to context. However, the personal rhythm, i.e. speaker-dependent duration, is not sufficiently emphasized, but it is an important factor for personal speech style, especially in stand-up comedy scenario. Other works \cite{explicit1-ada3, explicit2, explicit3, explicit4} train models with fillers to insert fillers (e.g. ``uh", ``um") in the text or speech. However, the fillers they choose are fixed and common for all speakers, without taking personal fillers and speaker-dependent pronunciations of fillers into account.

\begin{table*}[t!]
    \setlength{\abovecaptionskip}{0.1cm}
  \caption{The statistics of our stand-up comedy speech dataset}
  \renewcommand\arraystretch{1.0}
  \label{tab:statistics of dataset}
  \centering
  \setlength{\tabcolsep}{1.2mm}{
  \begin{tabular}{cccccccccc}
    \toprule
    \textbf{Comedian}  & \textbf{Gender} & \textbf{Clips} & \textbf{\makecell[cc]{Total \\ Duration(s)}} & \textbf{Words}  & \textbf{\makecell[cc]{Words \\ per Second}}  & \textbf{\makecell[cc]{Average Duration \\ of Pause(ms)}} & \textbf{\makecell[cc]{Average \\Energy}} & \textbf{\makecell[cc]{Average \\Pitch}} & \textbf{\makecell[cc]{Phoneme of \\ PF} }  \\
    \midrule
    A  &  male    & 120  & 505  &  3,357  &  6.65  & 206.1 & 0.24 & 0.19 & ``er2"  \\    
    B  &  female  & 140  & 622  &  3,824  &  6.15  & 188.3 & 0.11 & 0.43 & ``n i3 zh ii1 d ao4 b a5"  \\
    C  &  male    & 120  & 626  &  2,116  &  3.38  &575.5 & -0.18 & -1.04 &  None  \\
    D  &  female  & 120  & 587  &  2,891  &  4.93  & 142.3 & -0.11 & 0.12 & None  \\
    \bottomrule
  \end{tabular}}
   \vspace{-1.1em}
\end{table*}

\indent
Based on the above analysis, in this paper, we develop ComedicSpeech, a TTS system specially designed for stand-up comedy synthesis. We improve the personal speech style in three aspects. First, we model diverse personal prosody with a well-designed prosody encoder and employ conditional layer normalization to integrate personal prosody into the TTS model. Second, we enhance personal rhythm by learning speaker-dependent phoneme duration distribution with the conditional duration predictor. Third, we apply special tokens to represent personal fillers, imitating personal speaking habits in stand-up comedies.
Due to the lack of real-world stand-up comedy datasets, we build a multi-speaker stand-up comedy dataset. Although each speaker has about ten minutes for training, it is a good start for the stand-up comedy synthesis task. Experiment results also indicate that in such a low-resource scenario ComedicSpeech achieves better voice quality, voice similarity, and expressiveness in terms of MOS (Mean Opinion Score), SMOS (Similarity MOS), and Style MOS than baselines.


\section{Real-world Stand-up Comedy Dataset}
Some previous works build spontaneous datasets, but they mainly focus on conversations~\cite{explicit1-ada3, dataset1-dailytalk, dataset2-strategy}. In this paper, we construct a new real-world stand-up comedy dataset to support research on this new synthesis task by the following steps:
\begin{enumerate}
\item Raw Data Acquisition. We collect data from a Mandarin talk show called ``Rock $\&$ Roast" and select four stand-up comedians (two females and two males) with diverse personal styles as training speakers. We collect about 60-minute speeches for each as raw data.  

\item Data Denoising. Stand-up comedy speeches are mixed with noise, e.g., audience laughter and applause, which must be removed in order to train a high quality TTS model. We use Demucs~\cite{rouard2022hybrid, defossez2021hybrid}, a state-of-the-art music source separation tool to separate the human voice from the background music and other noise in our raw data.

\item Dataset Segmentation. We use an audio processing tool called Audacity \footnote{https://www.audacityteam.org/} to cut the audio. The data is sliced into segments of about 3$\sim$8 seconds.

\item Special Token Replacement. To better model the PFs of different comedians, we replace PFs in the data with special tokens and add them into the phoneme set. Table \ref{tab: an example of replacement} shows an example of replacing ``you know" by ``\textit{$\left \langle spc1 \right \rangle$}''. Please note that when two comedians have the same PF such as ``you know", we still use two different special tokens to replace them because they are pronounced in different styles. 
\end{enumerate}

The analyses and statistics results are shown in Table ~\ref{tab:statistics of dataset}. The statistics support our assumption that comedians have different prosody, rhythm, and fillers, which are part of their personal speech styles. First, the speech speed varies greatly. Among our four comedians, the first two are as fast as 6 words per second, while the other two are slower by 1-2 words per second. Second, the ``Average duration of pause" varies widely in comedians. Comedian C takes 575 ms as a pause, which is about two times longer than the other three comedians. The above two points show that the comedians vary widely in rhythm. The ``Average
Energy” and ``Average Pitch” indicate that the comedians vary widely in prosody. In addition, the four comedians use different PFs, as Table~\ref{tab:statistics of dataset} shows. Simply inserting fixed fillers like ``um" and ``uh" as in the previous work is not enough for our scenario. In summary, stand-up comedians have dramatic differences in personal prosody, rhythm, and fillers.

\section{Proposed Method}

\begin{figure*}[t!]
  \centering
  \includegraphics[width=0.85\linewidth]{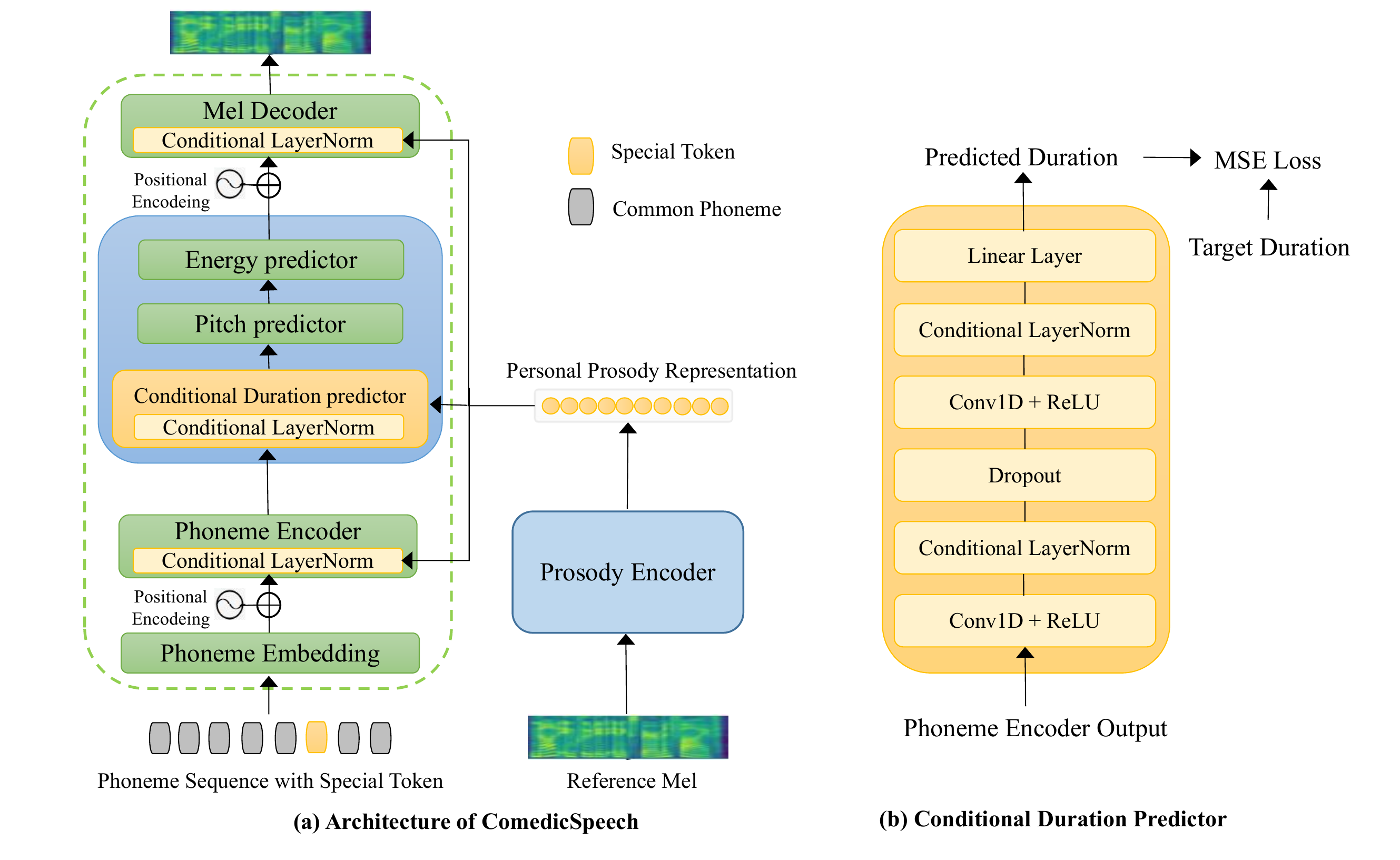}
  \caption{Model structure of ComdicSpeech, a multi-speaker TTS model based on the backbone of FastSpeech 2. }  \label{figure1:structure of ComdicSpeech}
  \vspace{-10pt}
\end{figure*}

\indent As shown in Figure ~\ref{figure1:structure of ComdicSpeech}, the overall architecture of ComedicSpeech is based on FastSpeech 2. We will describe our ideas and technique details in this section.

\subsection{Modeling Personal Prosody}
\indent Since comedians always show diverse personal prosody in performance, modeling and integrating their personal prosody is very important. Inspired by GST-Tacotron~\cite{implicit1-gst}, we apply a prosody encoder with GST-like architecture. The prosody encoder in ComidicSpeech extracts prosody representation through a weighted sampling of embedding from the shared prosody space. First, we map reference speech into a vector $R$, and then take it as the query vector used in Multi-head attention module. Define $P$ with dimension $d$ to be the basis prosody vector in the space shared by all speakers, then we obtain the prosody representation $E$:
\begin{equation}
\begin{split}
Q =& RW^{Q}, K = PW^{Q}, V = PW^{V},\\
&E=softmax(\frac{QK^{T}}{\sqrt{d}})V,
\end{split}
\vspace{-3.0em}
\end{equation}
where $W^{Q},W^{K},W^{V}$ are three trainable matrices, and $P$ is initialized randomly.
Then, following AdaSpeech 4~\cite{adaspeech4}, we integrate the personal prosody information into the TTS model by applying the conditional layer normalization (CLN) in encoder and decoder (as shown in Figure~\ref{figure1:structure of ComdicSpeech}(a) ). 
Specifically, the extracted prosody representation $E$ is used as the input of CLN, determining the scale and bias vector of FFT Block. 
Consequently, the prosody representation is flexible to change corresponding to different input audio.

\subsection{Modeling Personal Rhythm}

 The rhythm, which can be affected by phoneme duration, is one of the most important attributes of personal style, especially in stand-up comedy scenarios. To better model the personal rhythm and learn speaker-dependent duration distribution, ComedicSpeech has two improvements: 
 \begin{itemize}
 
 \item As shown in Figure~\ref{figure1:structure of ComdicSpeech}(b), we build a speaker-dependent duration predictor, which introduces more personal rhythm information to TTS model. Specifically, we use the conditional layer normalization in the duration predictor, generating adaptive scale vector $\gamma$ and bias vector $\beta$ in the duration predictor as follows:
\begin{equation}
\gamma = W^{\gamma}E ,\ \beta = W^{\beta}E ,
\end{equation}
where $W^{\gamma}$ and $W^{\beta}$ are linear layers for scale and bias respectively.
\item We increase penalties for inaccurate duration predictions by directly calculating MSE loss between the target duration and the predicted duration (See Figure~\ref{figure1:structure of ComdicSpeech}(b)), instead of calculating loss after converting them into logarithmic domain, like FastSpeech and AdaSpeech series.
The duration loss in ComedicSpeech can be calculated as follows:
\begin{equation}
\mathcal{L}_{duration} = \frac{1}{N}\sum_{i=1}^{N} (d_{target} - d_{predicted})^{2},
\end{equation}
where $d_{target}$ and $d_{prediction}$ represent target duration and predicted duration respectively.
\end{itemize}

\subsection{Modeling Personal Fillers}
\indent Special pronunciation of PF conveys a strong personal style. The PF varies from person to person, and even the same PF's pronunciation is different across people. Take comedian B for example, her PF ``you know'' usually has short duration less than 18 frames and ambiguous pronunciation caused by linking; whereas, when spoken by comedian A, C or D, ``you know'' lasts longer than 30 frames and is spoken clearly without linking. 
Therefore, different from AdaSpeech 3~\cite{explicit1-ada3} predicting the position of fixed PF (e.g. ``uh", ``um"), we introduce a special token to replace speaker-dependent PF (e.g. ``you know", ``I mean"). We find that learning PF as a token can ensure the model to learn speaker-dependent pronunciation of PF, instead of common pronunciation shared by all people. Also, it further reinforces the personal style in the stand-up comedy scenario.

\section{Experiments}
\subsection{Dataset}
Due to the lack of stand-up comedy dataset, first we pretrain ComedicSpeech on AISHELL-3 \cite{dataset3-AISHELL3}, a large-scale and high-fidelity multi-speaker Mandarin speech corpus. Then the model is finetuned on our stand-up comedy dataset. 
We split our dataset into training, validation, and test sets. All clips in test set are unseen during training. Finally we randomly choose clips synthesized by test set for human evaluation. The way of preprocessing on the speech and text data follows Adaspeech~\cite{Adaspeech}.   

\subsection{Implementation Details} 
Our model backbone is based on FastSpeech 2~\cite{fastspeech}, with the configurations follow it unless otherwise stated. Following \cite{implicit1-gst}, the prosody encoder contains three modules, a 6-layer convolutional network with stride 2 × 2 used to down sample reference mel, a GRU with hidden size 128 used to generate a query vector, and a prosody space with 8 prosody embeddings of 256 dimension to represent prosody by a weighted combination. 

\begin{table}[t!]
\setlength{\abovecaptionskip}{0.1cm}
\footnotesize
  \caption{The MOS (Mean Opinion Score), Similarity MOS, and Style MOS scores with 95\% confidence.}
  \label{tab:result}
  \setlength{\tabcolsep}{1mm}
  \centering
  \begin{tabular}{l|c|c|c}
    \toprule
      & \textbf{MOS$ \ (\uparrow )$} & \textbf{\makecell[cc]{Similarty \\ MOS} $ \ (\uparrow)$} & \textbf{\makecell[cc]{Style \\ MOS}$ \ (\uparrow )$} \\ 
    \midrule
    GT & $ 4.45 \pm 0.07 $  & $ 4.45 \pm 0.07 $ & $4.54 \pm 0.07 $ \\
    GT mel + Voc & $4.27 \pm 0.07$  & $4.28 \pm 0.09$ & $4.35 \pm 0.08$ \\
    \midrule
    FS 2 + spk emb & $2.74 \pm 0.12 $  & $ 2.47 \pm 0.13 $& $2.28 \pm 0.13 $  \\
    FS 2 + ref enc & $ 2.96 \pm 0.11 $ & $ 2.89 \pm 0.11 $ & $ 2.66 \pm 0.12 $  \\
    \midrule
    ComedicSpeech & \bm{$3.24 \pm 0.10 $}  & \bm{$2.90 \pm 0.11 $}  & \bm{ $ 2.80 \pm 0.12$ }  \\
    \bottomrule
  \end{tabular}
  \vspace{-2.5em}
\end{table}

\begin{figure*}[t!]
\centering 
\setlength{\belowcaptionskip}{-3mm}
\includegraphics[width=0.80\linewidth]{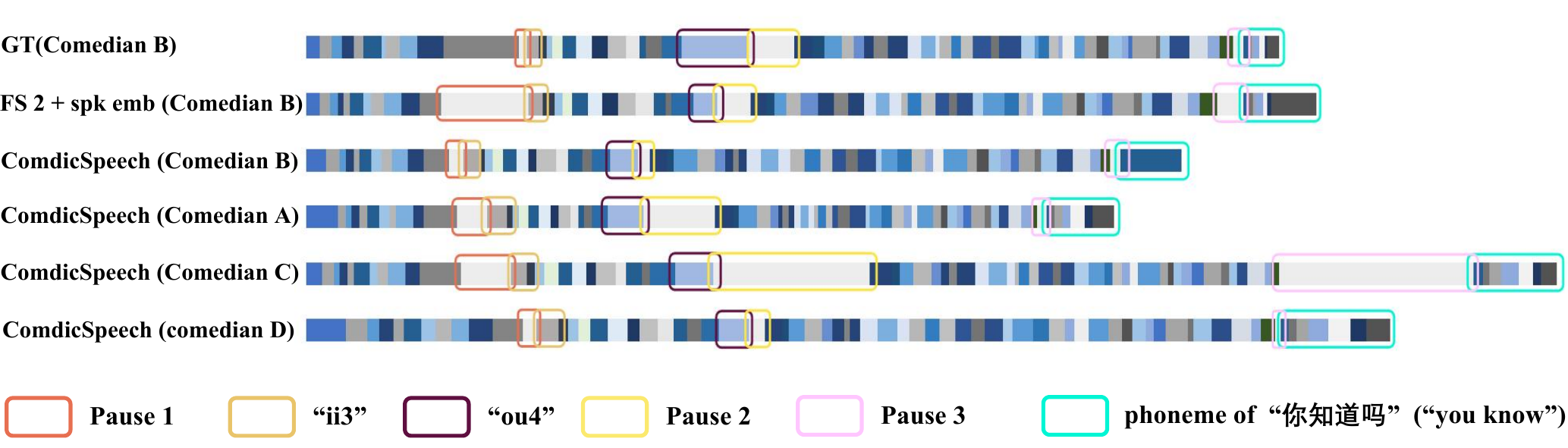}
\caption{Phoneme duration of the synthesized speech using the text content in Table~\ref{tab: an example of replacement}. A color segment represents a phoneme, the same color means the same phoneme, and the length of the segment represents the duration of the phoneme.}
\label{figure:duration}
\end{figure*}\vspace{-10pt}

\setParDis
We first pretrain ComedicSpeech for 300,000 steps on AISHELL-3 dataset~\cite{dataset3-AISHELL3}. Then we finetune the model on our constructed dataset for 100,000 steps. Both in training and finetuning stage, the reference speech fed in the prosody encoder is the ground truth mel-spectrogram extracted from the recording. And the prosody encoder is trained by all the losses in the model (e.g. duration loss, pitch loss) end-to-end. During inference, the reference audio is randomly selected from the training set.
In addition, we find that the second half of Fastspeech 2 synthesized speeches often have low voice qualities and electric sounds, so we added weights for the loss of the second half:
\begin{equation}
    Loss = loss_{first \: half} + \alpha \times loss_{second \: half}
\end{equation}

\setParDef

\subsection{Comparing ComedicSpeech with Baselines}
We conduct subjective listening tests including MOS (Mean Opinion Score), SMOS (Similarity MOS), and Style MOS to compare ComedicSpeech with baselines. Each clip is listened by 10 native volunteers. All volunteers are required to focus on voice quality, voice similarity, and expressiveness respectively when scoring MOS, SMOS, and Style MOS. We compare the synthesized speech of ComedicSpeech with several baselines: 1) GT, the ground truth recordings; 2) GT mel + Voc, using the ground-truth mel-spectrogram to synthesize waveform with HiFi-GAN vocoder.
3) FS 2 + spk emb, a FastSpeech 2 based multi-speaker TTS model using speaker ID to capture speaker-level acoustic conditions. 4) FS 2 + ref enc, a FastSpeech 2 based multi-speaker TTS model using a prosody encoder to extract speaker-level acoustic conditions. All settings use HiFi-GAN as the vocoder. Results are shown in Table \ref{tab:result}. 
It indicates that ComedicSpeech makes considerable improvements in terms of voice quality, voice similarity, and expressiveness in stand-up comedy scenario. 
Also, it should be mentioned that there still is a performance gap between ComedicSpeech and recording, because of the difficulty of stand-up comedy synthesis in low-resource scenario.

\begin{table}[t!]
\footnotesize
  \setlength{\abovecaptionskip}{0.1cm}
  \caption{The CMOS (comparison MOS), Similarity CMOS  and Style CMOS scores for ablation study. In this table, CLN stands for conditional layer normalization}
  \label{tab:ablation_smos_cmos}
  \centering
  \setlength{\tabcolsep}{1mm}{
  \begin{tabular}{l|c|c|c|c}
    \toprule
    \textbf{ID} & \makecell[c]{\textbf{Settings}} & \textbf{CMOS} & \textbf{\makecell[cc]{Similarity \\ CMOS}} & \textbf{\makecell[cc]{Style \\ CMOS}} \\ 
    \midrule
    \#1 & ComedicSpeech & \bm{$-$}  &  \bm{$-$} &  \bm{$-$}\\
    \midrule
    \#2 & \#1 $-$ duration CLN & $-\ 0.05$ & $-\ 0.23$ & $-\ 0.57$ \\
    \#3 & \#1 $+$ pitch and energy CLN & $-\ 0.65$ & $-\ 0.4$  & $-\ 0.86$\\
     \midrule
    \#4 & \#1 $-$ special token & $-\ 0.12$ & $-\ 0.02$ & $-\ 0.18$\\
    \bottomrule
  \end{tabular}}
  \vspace{-2.0em}
\end{table}

\subsection{Case Study}
We use the case from Comedian B in Table~\ref{tab: an example of replacement} to study whether our method can capture the personal style and differences across comedians. We draw lines of phoneme duration output by different models in Figure~\ref{figure:duration}, where the same color corresponds to the same phoneme. First, compared to the ground truth, our ComedicSpeech model for speaker B can better capture relative duration of phonemes, e.g., ``Pause 1'' and ``ii3'', and ``ou4'' and ``Pause 2'', than the baseline model FS 2 + spk emb~\footnote{Baseline FS 2 + spk emb performs better than FS 2 + ref enc in terms of duration loss, so we choose it to do the case study.}. That may result in better MOS, SMOS, and Style MOS scores. There is still room to improve on the absolute duration. Second, when using the ComedicSpeech models of other comedians, the same sentences have diverging speed of speech, pause duration, and PF duration. For example, Comedian C has the longest pause, i.e. ``Pause 2'', which is consistent with statistics in Table~\ref{tab:statistics of dataset}. He likes to pause for longer time and speak slowly. In addition, when looking close to the duration of PF, we find that ``you know'' is a PF for Comedian B but not for other comedians. Thus the other models do not take it as a whole special token. Consequently, they predicate longer duration for ``you know'', rather than our model for Comedian B, which makes sense.

    

\subsection{Ablation Studies}
We conduct a series of ablation experiments to verify the effectiveness of each design in ComedicSpeech, including conditional layer normalization in the duration predictor, and the special tokens. We evaluate the CMOS (comparison MOS, focus on voice quality), Similarity CMOS (focus on voice similarity), and Style CMOS (focus on expressiveness) of the synthesized speech when removing each component in ComedicSpeech. The results are shown in Table \ref{tab:ablation_smos_cmos}. 

We find that removing conditional layer normalization (CLN) from duration predictor impairs voice quality, voice similarity and speech expressiveness. However, adding CLN to all variance adaptor (e.g. pitch predictor, energy predictor) does not bring significant gain but leads to a drop of all three metrics. This is because CLNs in encoder and decoder are sufficient for the modeling of pitch and energy, and more CLNs in variance adaptor cause the model to focus too much on comedian’s accent, creating strange tunes. However, the modeling of duration is so important and difficult in this scenario that only two CLNs cannot model it well. So our adding CLNs in only phoneme encoder, duration predictor, and mel decoder makes sence.
In addition, directly regarding PF as an ordinary phoneme sequence impairs voice quality, voice similarity, and speech expressiveness, causing a decline in all three metrics. Therefore, our using a special token to replace the whole PF does work.

\section{Conclusions}
In this paper, we propose ComedicSpeech, a TTS model specially designed for stand-up comedy synthesis in low-resource scenario. We improve the personal speech style in three aspects. First, we introduce the prosody encoder to extract personal prosody of different comedians. Then we learn a more accurate duration distribution to improve personal rhythm by duration predictor. Next, we model the personal fillers by introducing a special token to ensure that model learns speaker-dependent pronunciation. Due to the lack of dataset, we build a multi-speaker dataset from a real-world stand-up comedy scenario. Experimental results show that our proposed ideas improve voice quality, voice similarity, and expressiveness for different comedians over strong baselines. 
In future, we will improve the modeling ability of the diverse personal style and speech expressiveness in stand-up comedy scenario. We will also try to combine TTS model with pretrain language model, to insert personal fillers for comedians automatically.




\newpage

\bibliographystyle{IEEEtran}
\bibliography{main}

\end{document}